\documentstyle[epsfig]{article}
\begin{document}

%------------------begin definitions--------------------------------
\def   \ni {\noindent}
\def   \cl {\centerline}
\def   \vs {\vskip}
\def   \hs {\hskip}
 
\def   \ssk {\vskip  5truept}
\def   \sk  {\vskip 10truept}
\def   \bsk {\vskip 15truept}
 
\def   \newpage {\vfill\eject}
\def   \newline {\hfil\break}
 
\def   \u {\underbar}

 %------------end definitions-----------------------------------------
%

\hsize 5truein
\vsize 8truein
\font\abstract=cmr8
\font\keywords=cmr8
\font\caption=cmr8
\font\references=cmr8
\font\text=cmr10
\font\affiliation=cmssi10
\font\author=cmss10
\font\mc=cmss8
\font\title=cmssbx10 scaled\magstep2
\font\alcit=cmti7 scaled\magstephalf
\font\alcin=cmr6 
\font\ita=cmti8
\font\mma=cmr8
\def\ref{\par\noindent\hangindent 15pt}
\null
%\vskip 3.0truecm
%\baselineskip = 12pt

% beginning of font "title"

\title{\ni Creation of a dense torus in the coalescence of a black hole
with a neutron star
}                                               
% beginning of font "author and affiliation"
\bsk \bsk
\author{\ni W{\l}odzimierz Klu\'zniak$^{1,2}$ and William H. Lee$^{1,3}$}
                                                      
\bsk
\affiliation{$^1$University of Wisconsin, Physics Dept., Madison WI 53706, USA}

\affiliation{$^2$Nicolaus Copernicus Astronomical Center, Bartycka 18,
00-716 Warszawa, Poland
}

\affiliation{$^3$ Universidad Nacional Aut\'{o}noma de M\'{e}xico,
 Apdo. Postal 70-264,
  04510 M\'{e}xico 
 }                                              
\bsk
\baselineskip = 12pt
% beginning of font "abstract and keywords"
\abstract{ABSTRACT \ni
We used a newtonian SPH (smooth-particle hydrodynamics) code
 to follow the final stages of evolution of a coalescing binary system
 of a neutron star and a black hole. We find that the outcome of the
 ``merger'' is very sensitive to the equation of state describing the
 neutron star.  A neutron star with a soft equation of state (polytrope
 with index $\Gamma=5/3$) is completely disrupted, and a fairly large
 and long-lived accretion torus is formed.
 }
\bsk
\baselineskip = 12pt
\keywords{\ni KEYWORDS:  neutron stars -- black holes -- binaries: evolution --
 hydrodynamics -- gamma-ray bursts}               
\bsk
\baselineskip = 12pt
% beginning of font "text"
\text{\ni 1. INTRODUCTION}
\ssk
\ni
  
The coalescence of a stellar-mass black hole with a neutron star
has been discussed as a possible site of the {\it r\/}-process and
a possible source of gamma-ray bursts (Lattimer \& Schramm 1976).
It has been expected  that in such an encounter, when the loss of
angular momentum to gravitational radiation brings the commponents
of the binary sufficiently close,  tidal forces will
disrupt the neutron star completely (Wheeler 1971). The resulting toroidal
structure was supposed to be sufficiently long-lived that its ultimate
accretion onto the black hole would proceed on the viscous timescale, thus
possibly allowing the appearance of a gamma-ray burst
 lasting several seconds
or minutes (Paczy\'nski 1991, M\'esz\'aros and Rees 1993).
\bsk
\ni 2. RESULTS FOR A SOFT E.O.S ($\Gamma=5/3$)\ssk
\ni 
We have performed two series of SPH runs, with about 17 000 particles each,
in which the only essential difference was the polytropic index
of the e.o.s. used to model the initial neutron star. The results for
$\Gamma=3$ are as reported previously (Klu\'zniak and Lee 1998).
But for $\Gamma=5/3$,
we find that the star is completely disrupted. Here we exhibit
representative results obtained for an initial mass ratio of
$q=0.31$. In the calculation,
the mass of the neutron star
is $M_*=1.4M_\odot$ and its (unperturbed) radius $R_*=13.4\,$km.
%--------------------------  figure 1
%this section shows how to insert a figure in the text
\begin{figure}%[t]
\centerline{\psfig{file=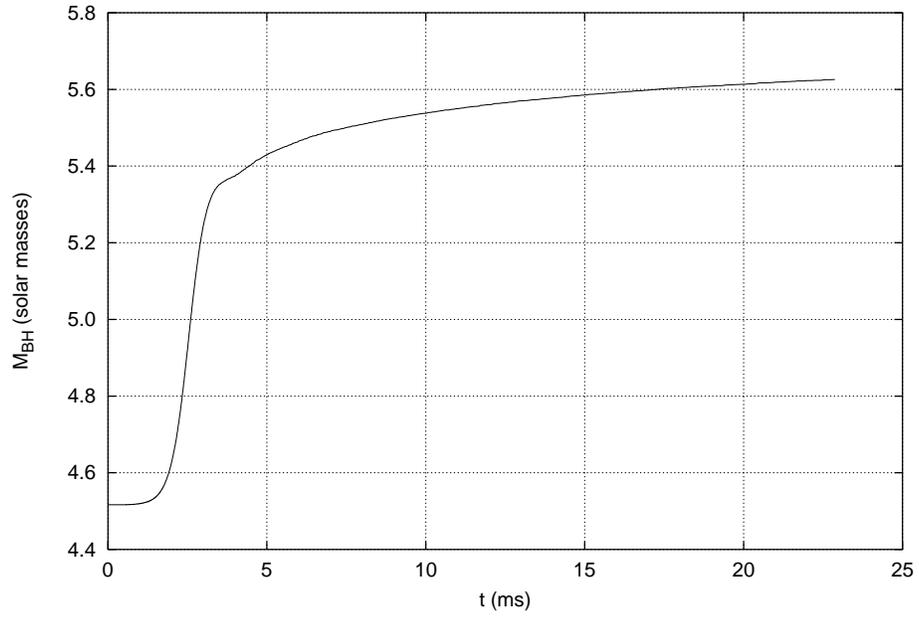, angle=0, width=\textwidth}}
\caption{FIGURE 1. Black hole mass as a function of time.}
\end{figure}

%---------------------------------
%--------------------------  figure 
%this section shows how to insert a figure in the text
\begin{figure}%[t]
\centerline{\psfig{file=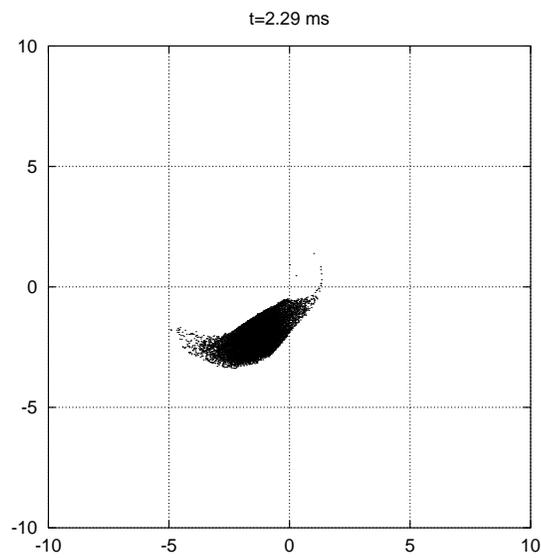, angle=0, width=3truein}}
\caption{FIGURE 2. Positions of the SPH particles projected onto
the orbital plane at t=2.29 ms. The outline of the black hole is faintly
visible, mostly in the first quadrant.}
\end{figure}

%---------------------------------

%--------------------------  figure 3a
\begin{figure}%[t]
\centerline{\psfig{file=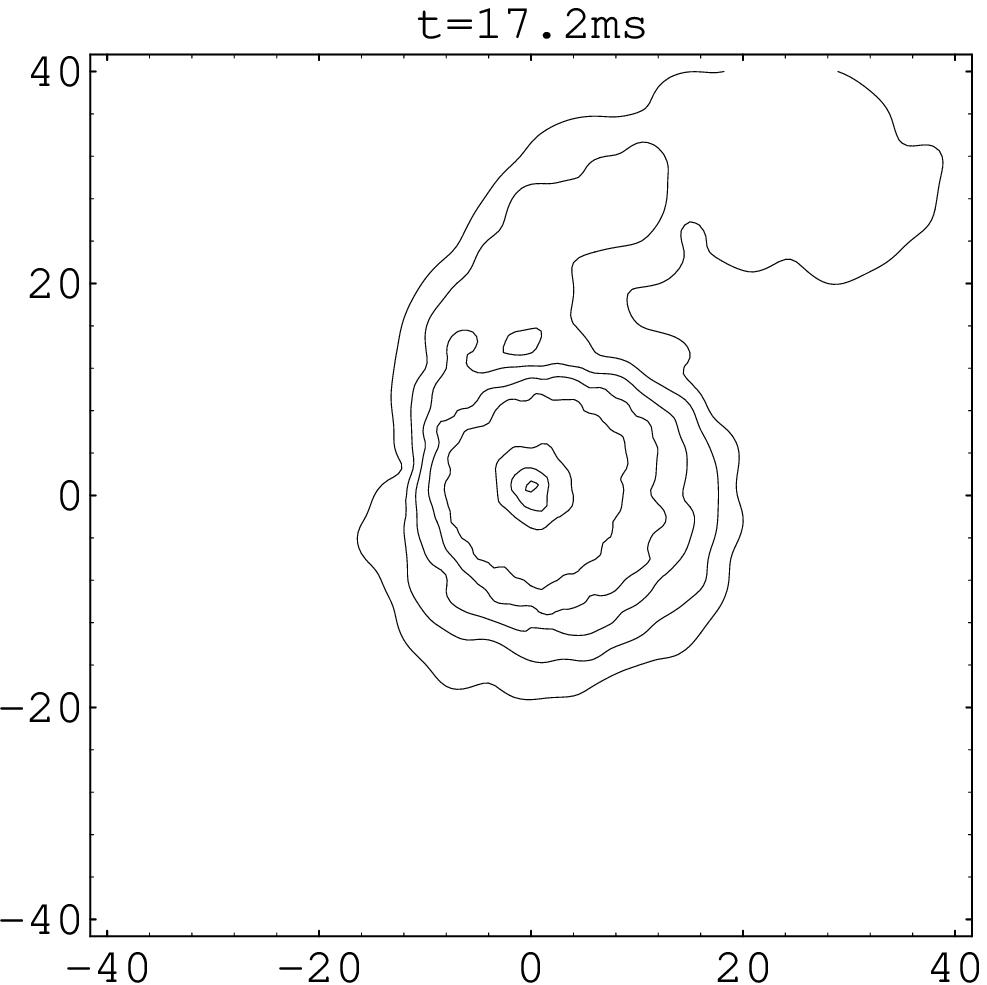, angle=0, width=3truein}}
\caption{}
\end{figure}

%---------------------------------
%--------------------------  figure 3b
\begin{figure}%[t]
\centerline{\psfig{file=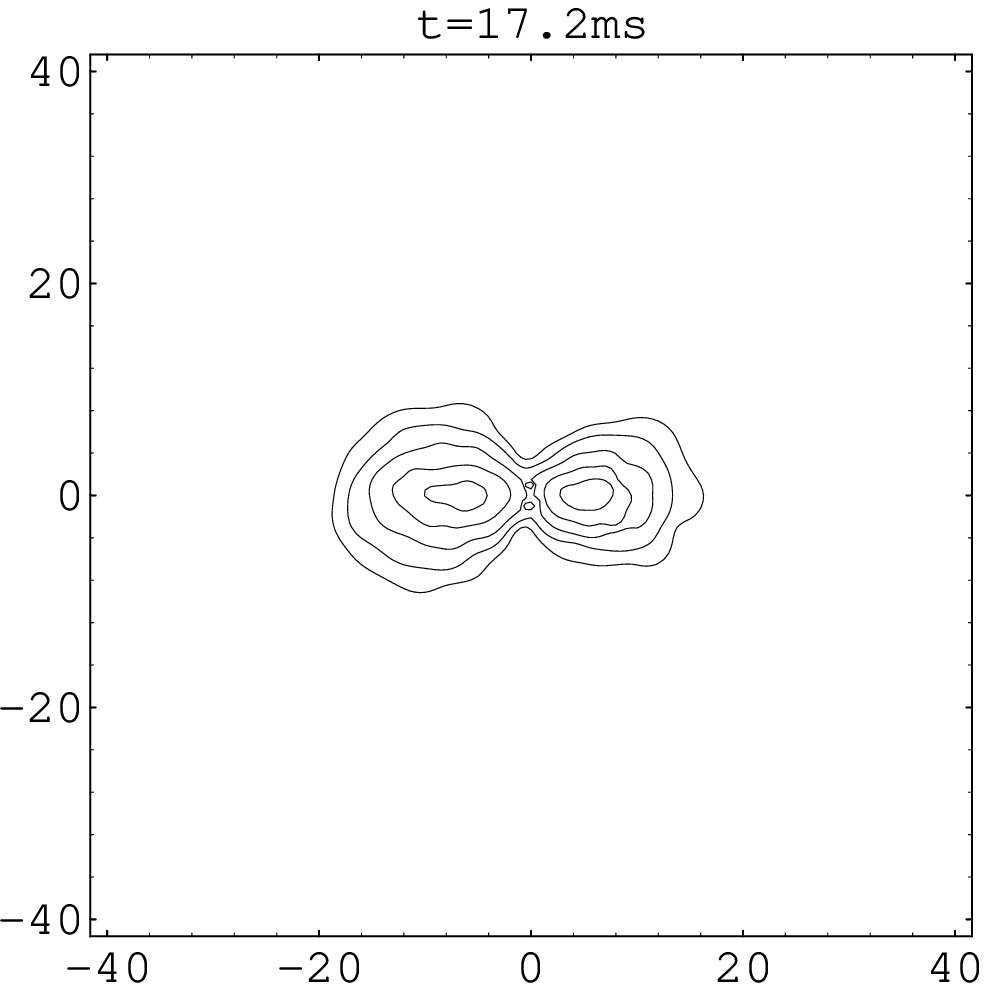, angle=0, width=3truein}}
\caption{FIGURE 3. Logarithmic density contours in the orbital plane
(top panel) and in a meridional plane (bottom panel) at t=17.2 ms, after
the rapid phase of accretion is over.}
\end{figure}

%---------------------------------

The results presented were obtained for an initially tidally locked
binary. The initial conditions are obtained by relaxing the polytrope
in the presence of the tidal field, as desribed elsewhere
(Rasio \& Shapiro 1992, Lee \& Klu\'zniak 1995, 1998). For a full
description of the code see Lee 1998. Upon relaxing the polytrope we
set time equal to zero and allowed the system to evolve dynamically in
the presence of a potential mocking up gravitational radiation
reaction for two point masses orbiting each other.
We have turned off this potential at time t=2.865 ms
% (25 in units of $\tilde t = \sqrt{R_*^3/GM_*}= 0.1146\,$ms),
when the star was being clearly
disrupted; the subsequent computation was purely newtonian.
The black hole was modeled as a newtonian point-mass absorbing matter
(and momentum) at a spherical boundary of Schwarzschild radius.

In Fig.~2
we show the positions of the SPH particles (projected onto the orbital
plane) at time 
$t=2.29\,$ms when the black hole is
entering a phase of rapid accretion (Fig. 1). Note that $1.1M_\odot$
is accreted in about 15 ms.

Fig.~3 shows the distribution of matter at
$t=17.2\,$ms, after the black hole is no longer in
the phase of rapid accretion.The lowest contour corresponds
to the value of density
$\rho=10^{-6}M_*/R_*^3=1.14\times10^{12}$kg/m$^3$, distance is in units of the
unperturbed stellar radius ($R_*=13.4\,$km).
Inspection of the density
contours (logarithmically spaced every one-half decade), shows that the
star has been completely disrupted and a large accretion torus has formed.
%
%\noindent

\bsk
\ni 3. DISCUSSION
\ssk
\ni

For the first time in numerical calculations, we have obtained a complete
disruption of the star and the persistence of a long-lived
toroidal structure
 as the outcome of the final stages of evolution of a
neutron star--black hole binary. Modeling the initial neutron star with a
relatively soft e.o.s.
% ($\Gamma=5/3$)
was the essential step in obtaining
this result in our newtonian simulations. But even in this case, only a small
fraction of the mass of the neutron star is locked up in the torus
(at $t=17.2\,$ms, $0.28 M_\odot$ is within $20R_*$ of the black hole).

\sk
\baselineskip = 12pt
{\abstract \ni ACKNOWLEDGMENTS
This work has been supported in part by KBN (grant P03D01311).

}

\sk
\baselineskip = 12pt

% beginning of font "references"

{\references \ni REFERENCES
%\ssk
\ref Klu\'zniak, W. Lee, W.H. 1998, ApJ 494, L53
\ref Lattimer, J.M., Schramm, D.N. 1976, ApJ 210, 549
\ref Lee, W.H. 1998, Ph.D. Thesis, University of Wisconsin
\ref Lee, W.H., Klu\'zniak, W. 1995, Acta Astron. 45, 705
\ref Lee, W.H., Klu\'zniak, W. 1998, ApJ submitted, astro-ph/9808185
\ref M\'esz\'aros, P., Rees, M.J. 1993, MNRAS 257, 29P
\ref Paczy\'nski, B. 1991, Acta Astron. 41, 257
\ref Rasio, F., Shapiro, S.L. 1992, ApJ 401, 226
\ref Wheeler, J.A. 1971, Pontificae Acad. Sci. Scripta Varia 35, 539
}
\end{document}